\documentclass[12pt]{article}
\usepackage{epsfig,cite,amsmath,amssymb,a4p}
\newlength{\picwi}
\usepackage[pagewise]{lineno}

\newlength{\twofig}  
\begin{document}
%
\def\jour#1#2#3#4{{#1} {\bf#2} (19#3) #4}
\def\jou2#1#2#3#4{{#1} {\bf#2} (20#3) #4}
\def\AP{Acta Phys. Pol. {B}}
\def\CP{Comp. Phys. Comm.}
\def\EPJ{Eur. Phys. J. {C}}
\def\IJ{Int. J. Mod. Phys. {A}}
\def\IAN{Izv. Akad. Nauk: Ser. Fiz.}  
\def\JL{JETP Lett.}
\def\JP{J. Phys. {G}}
\def\ML{Mod. Phys. Lett. {A}}
\def\NC{Nuovo Cim. {A}}
\def\NIM{Nucl. Instr. Meth. {A}}
\def\NPA{Nucl. Phys. {A}}
\def\NP{Nucl. Phys. {B}}
\def\NPBP{Nucl. Phys. {B} (Proc. Suppl.)}
\def\NPYF{Sov. J. Nucl. Phys.}
\def\Phy{Physica {A}}
\def\PL{Phys. Lett.  {B}}
\def\PRC{Phys. Rev. {C}}
\def\PRD{Phys. Rev. {D}}
\def\PRL{Phys. Rev. Lett.}
\def\PRp{Phys. Rep.}
\def\PPNP{Prog. Part. Nucl. Phys.}
\def\RPP{Rep. Prog. Phys.}
\def\Usp{Physics - Uspekhi}
\def\ZP{Z. Phys.  {C}}
%
%
\def\nopar{\noindent}
\newcommand{\bec}{Bose-Einstein Correlations}
\def\Ecm{\sqrt{s}}
\def\Ecmn{\sqrt{s_{\rm NN}}}
\def\Ecmp{\sqrt{s_{\rm pp}}}
\def\Ecme{\sqrt{s_{\rm ee}}}
\def\ep{$\rm{e}^+\rm{e}^-$}
\def\app{${\bar {\rm p}}{\rm p}$}
\def\timesplus{\times \hspace{-.3122cm}+}
\def\triangleinsquare{\square\hspace*{-.29cm} \scriptstyle \triangle}
\def\blktriangdowninsqr{ \square\hspace*{-.31cm} \blacktriangledown}
\def\blktrianginsqr{\square\hspace*{-.29cm} \blacktriangle}
\def\timesplussquared{{\scriptstyle \boxtimes} \hspace{-.26cm}+}
\def\vs{\vspace*}
\def\hs{\hspace*}
\def\bi{\bibitem}
\def\fig{Fig}
\def\sect{Sect.}
\def\col{Collaboration}
\def\ea{{\sl et al.}}
\def\eg{{\sl e.g.}}
\def\vrs{{\sl vs.}}
\def\ie{{\sl i.e.}}
\def\va{{\sl via}}
\def\ct{\cite}
\def\lss{\pm\,\pm}
\def\al{\langle}
\def\ar{\rangle}
\def\leq{\leqslant}   
\def\geq{\geqslant}
\def\lssim{\lesssim}
\def\gtsim{\gtrsim}
\def\pythia{{\sc Pythia}}
\begin{titlepage}
\begin{flushright}
{\large hep-ph/0410324}\\
{\large  CERN-PH-TH/2004-213}\\
\end{flushright}
\vspace*{.8cm}
\bigskip
 %
\begin{center}
\begin{boldmath}
\bf 
 \Large 
 On similarities of 
 bulk observables 
\\
in nuclear and particle collisions 
\end{boldmath}
\end{center}
 %
 %
\begin{center}
{\large 
Edward K.G. Sarkisyan$^{a,b}$ and  Alexander S. Sakharov$^{a,c,d}$\\
\bigskip
{ \footnotesize \it
$^a$ Department of Physics, CERN, CH-1211 Geneva 23, Switzerland \\
$^b$ Department of Physics, the University of Manchester, Manchester M13 
9PL, UK  \\
$^c$ Swiss Institute of Technology, ETH-Z\"urich, 8093 Z\"urich,
Switzerland \\
$^d$ INFN Laboratory Nazionali del Gran Sasso, SS. 17bis 67010 
Assergi (L'Aquila), Italy
\\
}
}
\end{center}
\bigskip
\bigskip
 %
\begin{abstract}
\nopar
 We study the regularities in the multiparticle production data
obtained
from different types of collisions  indicating the universality of the 
hadroproduction 
process.    
 The similarities of such bulk 
variables like the charged particle  mean multiplicity and 
the  pseudorapidity density at midrapidity measured
 in nucleus-nucleus, (anti)proton-proton and \ep\ interactions are
analysed 
according to the dissipating energy of participants and their types. 
 This approach shows a  
good agreement 
with the measurements in a wide range of  nuclear collision 
energies from 
AGS to RHIC. The predictions  up to the LHC energies are made 
and compared to  
experimental extrapolations.  
 \end{abstract}
\vspace{1.cm}
\end{titlepage}

\newpage

{\bf 1.} Nucleus-nucleus collisions at RHIC probe matter at high 
densities and 
temperatures ever reached and provide us with an opportunity to 
investigate 
strong 
interactions 
up to extremely high energy density parton collisions. 
 Of fundamental interest are    
 bulk observables such as
 multiplicity and particle densities (spectra), which are  
 sensitive tools for probing the dynamics of strong interactions. 
  They give us information about the system formed in high energy 
collisions, both after   
cooling and 
hadronisation and as well 
during the 
formation 
and 
evolution of the collision initial state, and are thus powerful in 
distinguishing between different particle production models.
 Recent measurements at RHIC revealed striking evidences in the hadron 
production process including
 some universality between  such  basic observables as the mean 
multiplicity and the midrapidity density
in complex  ultra-relativistic nucleus-nucleus collisions \vrs\ those 
measured 
in relatively ``elementary'' 
\ep\ interactions.
The values of these 
bulk observables
are 
found to be similar
 for  both types of reactions  
 when measurements are
 normalised to  pairs of participants 
(``wounded''  nucleons \ct{woundN} 
in  
heavy 
ion 
collisions) 
at the nucleon-nucleon center-of-mass (c.m.) 
energy, $\Ecmn$,  comparable to the c.m. energy $\Ecme$ of 
\ep\ annihilation \ct{ph-sim,ph-rev}. 
This phenomenon is found to be independent of the  
energy spanning  from $\Ecmn=$ 19.6 GeV to 200 GeV at RHIC.
 Assuming a universal mechanism of hadron production in both types of 
interactions which then depends only on the amount of energy 
transformed into 
particles produced, one would expect the same value of the observables to 
be 
obtained in hadron-hadron collisions 
 at close c.m. energies. 
 However, this is not the case: comparing measurements in hadronic data 
 \ct{ua5-53900,cdf} to
the findings at RHIC, 
one obtains 
\ct{ph-sim,ph-rev,ph-56130,phxAph,st02,busza,phx-syststuds,phx-milov,phx-rev,br-rev,ph-talk04}
quite lower values in hadron-hadron collisions.
In the meantime, recent RHIC data from deuteron-gold 
interactions at 
$\Ecmn=$ 200~GeV unambiguously points to the same values of the mean 
multiplicity as measured in antiproton-proton collisions 
\ct{ph-rev,busza,ph-talk04,ph-dAu-talk,ph-dAu}.

In this paper, we give a phenomenological interpretation of the above 
regularities obtained 
and compare these to the earlier data at lower energies. 
 The particle production is considered to depend on the amount of the 
participant
energy 
dissipated in collision. 
 The comparison between different reactions is made on the basis of the 
type of participating patterns.
 This consideration introduces a tripling factor to be taken into 
account for the c.m. energy 
of a reaction and for appropriate treatment of participants to  correctly 
compare the bulk variables from different reactions.
Good agreement between 
the description proposed here
and the experimental measurements is found. 
 \medskip

{\bf 2.} In the consideration given here,
the whole
process of a collision is interpreted as the expansion and break-up
into particles of an initial state, in which the whole available energy is
assumed to be concentrated in a small Lorentz-contracted volume.
 There are no any restrictions 
due to 
the conservation of quantum numbers besides energy 
and
momentum constraints allowing therefore to link the amount of energy 
deposited in 
the collision zone and features of bulk variables in different reactions. 
 This approach resembles the Landau
phenomenological hydrodynamical description of multiparticle production 
in relativistic particle collisions \cite{landau}. 
 Though the hydrodynamical description does not 
match ideally the data on multiparticle production in the whole range of 
pseudorapidity,  the model-based
  energy-scaling 
law of multiplicity
 gives good agreement with the measurements 
 in
 such different reactions as nucleus-nucleus, pp,
\ep\ and $\nu$p collisions \ct{feinberg1,landau-exp}.  
 Recently, 
the Landau model prediction for the longitudinal pion transport has been 
observed 
to  be
well 
within 
5\% accuracy with the heavy-ion data 
\ct{br-meson}. 
 This indicates that the main asserts of the Landau approach are useful 
to estimate fractions of the energy dissipated into particles produced in 
different reactions, particularly in nucleus-nucleus collisions 
\ct{bjorken}. 

 As soon as the collision of two Lorentz-contracted
particles leads to the full thermalisation of the system before extension, 
one
can assume that the production of secondaries is defined by the fraction
of energy of participants deposited in the volume of thermalized system at 
the collision
moment. This implies that there is a difference between results of 
collisions of structureless particles like electron and positron and 
composite 
particles
like proton, the latter considered to be built of  
constituents. 
 Indeed, in composite particle collisions
not all the constituents deposit their energy when they form a 
small
Lorentz-contracted volume of thermalised initial state.
As a result, the leading particles 
\ct{leadp} formed out 
of
those constituents which are not trapped in the interaction volume, carry
away a part of energy effectively making it unavailable to participate in
production of secondaries. From the other side, colliding structureless
particles are ultimately stopped as a whole in the initial state of
thermalized collision zone depositing their total energy in the small
Lorentz-contracted volume. The latter makes this energy of the
incoming particles wholly available for the production of secondaries.

A single nucleon represents a superposition of three constituent
quarks. In this additive quark picture \ct{constq}, 
most often only one quark from each nucleon 
contributes to the interaction, while other
quarks are considered to be spectators.  Thus the initial thermalized 
state which is
responsible for the number of produced secondary particles is pumped in
only by the energy of the interacting single quark pair. The quark 
spectators 
form
hadrons being not stopped in the thermalized volume at the collision 
moment do not participate in secondary particle production.  The later 
means
that only 1/3 of the entire nucleon energy is available for particle
production. In  \ep\ annihilation, the incident particles are
structureless and therefore, the total interaction energy is deposited in 
the
thermalized collision zone. From this and the above consideration, 
one expects
that the resulting bulk variables like the multiplicity and rapidity
distributions should show identical features in proton-proton collisions
at the c.m. energy $\Ecmp$ and \ep\ interactions at the c.m. energy
$\Ecme \simeq\Ecmp/3$.
 Note that for the mean multiplicity, similar behaviour was obtained 
experimentally in 
the beginning of LEP activity \ct{ee3pp}. 

In heavy ion collisions, more than one quark
per nucleon interacts due to the large size of nucleus
and the long path of interactions inside the nucleus. The more central the 
nucleus-nucleus collision is, the more interactions occur and the larger
energy amount is spent for secondary particle production.
 In central nuclear collisions, a contribution of constituent quarks 
rather than participating nucleons
seem to determine the properties of produced particle distributions 
\ct{voloshin}.
In
the most central collisions, the density of matter is so 
high (almost saturated)  that all three constituent quarks 
from each nucleon may participate 
nearly simultaneously 
in collision 
depositing their
energy coherently into thermalized collision zone. 
The total entire energy
of nucleons included in the most central fraction of colliding nuclei is
available for secondary particle production. 
Recalling that in proton-proton collisions,  where only one out of three 
constituent quarks from each proton
interacts, one  
expects the features of the bulk variables per pair of participants 
measured in the 
most central 
heavy-ion 
interactions at the c.m. energy $\Ecmn$  to be similar to those  
from
proton-proton collisions at three times larger c.m. energy, $\Ecmp \simeq 
3\, \Ecmn$.
This makes the most central
collisions of nuclei akin to \ep\ collisions 
at the same c.m. energy
from the point of view
of the resulting bulk variables. 
 \medskip

\begin{figure}[t]
 \vspace*{-.5cm}
\epsfysize=14cm
  \begin{center} 
\epsffile{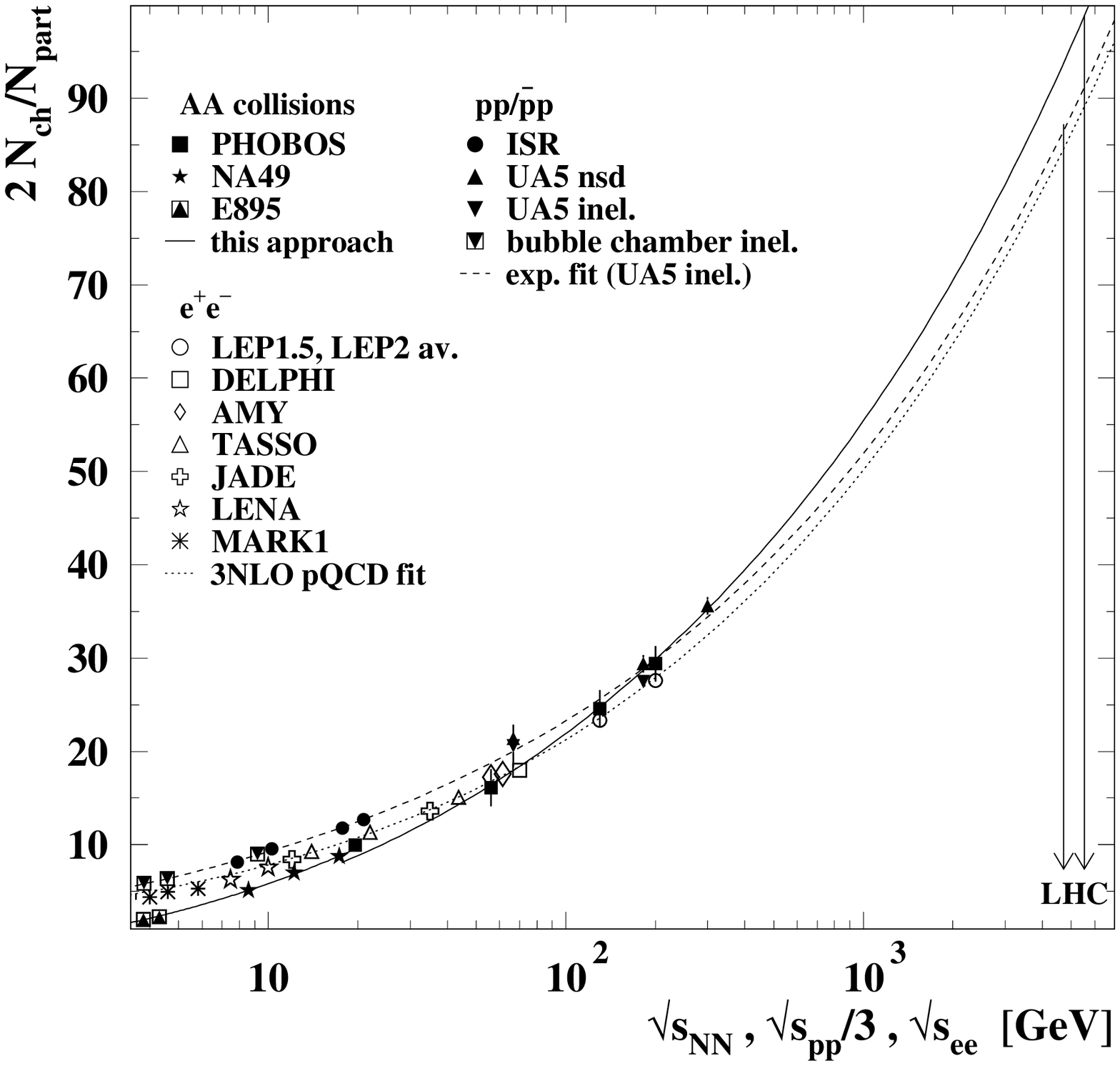}
  \end{center} 
 \vs{-.5cm}
\caption{
 The charged particle mean multiplicity per participant pair
as a function of the c.m. energy. 
 The solid and combined symbols show the multiplicity value 
 of the most central 
heavy-ion (AA) collisions at RHIC 
 as measured by PHOBOS \col\  
($\scriptstyle \blacksquare$)  
in \ct{ph-sim,ph-rev,ph-mult,ph-colgeom}, 
and by NA49 \col\ at CERN SPS \ct{na49-mult} 
($\scriptstyle \bigstar$) 
and by E895 \col\ at AGS \ct{agsmult} 
($\blktrianginsqr$)
(see also \ct{ph-sim}), 
 and, 
 in \app\ collisions, the  measurements made at CERN by  
 UA5 \col\ ($\blacktriangle$ for non-single diffractive, 
$\blacktriangledown$ for inelastic events) at 
$\Ecmp=$ 546 GeV
\ct{ua5-546} and $\Ecmp=$ 200 and 900 GeV \ct{ua5-zp} 
and, at lower c.m. energies,
 in pp collisions obtained at CERN-ISR 
($\bullet$) \ct{isr-thome} and from bubble chamber experiments 
\ct{bubblechamber,fnalmult} ($\blktriangdowninsqr$) , the latter 
compiled and analysed
in 
\ct{eddi}. 
 (The inelastic UA5 data at $\Ecmp=$~200~GeV is extrapolated in 
\ct{ph-dAu-talk} from the 
limited rapidity range to the full one.) 
 The open symbols show the \ep\ measurements:  the 
high-energy LEP 
mean 
multiplicities averaged here from the recent
data values  ($\scriptscriptstyle \bigcirc$) 
at 
 LEP1.5 $\Ecme =$~130~GeV in \ct{lep1.5o,lep1.5-2adl} and LEP2
$\Ecme
=$~200~GeV in 
\ct{lep1.5-2adl,lep2o}, and the lower-energy data as measured by DELPHI 
\ct{delphi70} ($\scriptstyle \square$), 
TASSO \ct{tasso} ($\scriptstyle \triangle$), 
AMY \ct{amy}  ($\scriptstyle \diamondsuit$), JADE \ct{jade} (+), LENA 
\ct{lena} 
($\star$), and MARK1 \ct{mark1} 
 ($\timesplus$) 
Collaborations. 
 (See also 
\ct{opalcomp,pdg,biebel} 
for data on \ep\ and pp/\app\ collisions).
 The solid line shows the calculations from Eq.~(\ref{prap0}) 
based on our approach and using the corresponding fits (see text).
 The dashed and dotted lines show the fit to the pp/\app\  data from 
\ct{ua5-546} 
and 
the 3NLO  perturbative QCD fit \ct{dremin} to \ep\ data 
 by ALEPH \ct{lep1.5-2adl}.
 The arrows show the expectations for the LHC.
 }
\label{fig:multshe}
\end{figure}

{\bf 3.} 
 Applying the above consideration, we compare in 
Fig.~\ref{fig:multshe}, 
in the  $\Ecmn$ energy range from a few GeV 
to 200~GeV, 
 the c.m. energy dependence of  
the mean multiplicity  as measured in nucleus-nucleus 
\ct{ph-sim,ph-rev,ph-mult,ph-colgeom,na49-mult,agsmult}, 
and \ep\ interactions 
\ct{lep1.5o,lep1.5-2adl,lep2o,delphi70,tasso,amy,jade,lena,mark1} \vrs\ 
that 
but at the three times larger c.m. 
energy 
in 
 (anti)proton-proton 
 pp/\app\
collisions \ct{ua5-546,ua5-zp,isr-thome,bubblechamber,fnalmult}.
 The multiplicity values in 
 nuclear 
collisions are given divided by the number of participant pairs and the 
energy 
value 
gives the c.m. energy per nucleon, $\Ecmn$. 
For the data on
\ep\ annihilation at energies above the Z$^0$ peak, we give 
the 
multiplicity 
values averaged\footnote{The averaging 
 procedure is adopted from the PDG review \ct{pdg} for the averaging 
over $N$ 
 measurements. 
 The errors are multiplyed by the $S$-factor, where $S= \sqrt{ 
 \chi^2/(N-1)} >1$ or $S=1$ otherwise.}   
here from those   
recently published by LEP experiments 
at LEP1.5 $\Ecme =$~130~GeV \ct{lep1.5o,lep1.5-2adl} and LEP2 
$\Ecme 
=$~200~GeV 
\ct{lep1.5-2adl,lep2o} energies: $23.35\pm 0.20 \pm 0.10$ (LEP1.5) and 
$27.62 \pm 0.11 \pm 0.16$ (LEP2). 
 Figure shows also the average multiplicity fit to pp/\app\   data 
obtained in \ct{ua5-546} and the ALEPH fit \ct{lep1.5-2adl}, 
based on the 3NLO  
perturbative QCD calculations \ct{dremin},
to \ep\ data.

First, one can 
see  from Fig.~\ref{fig:multshe} that the pp/\app\ data 
  are very 
close to 
the data from \ep\ annihilation measured at the c.m energy 
$\Ecme=\Ecmp/3$. 
This nearness, as the fits confirm    
spanning from  GeV to TeV c.m. energies,  
decreases the 
already small deficit in the \ep\ data compared to the pp/\app\ data 
as the 
energy increases. Note that at TeV energies this deviation is almost 
five-six times 
smaller the area covered by the fit errors.
 The deviation can be attributed to the inelasticity factor, or leading 
particle effect \ct{leadp} in pp/\app\ collisions, which is known to 
decrease 
with the 
c.m. energy. Then, at lower c.m. energies, some fraction of the energy of 
spectators contributes more into the formation of the initial state, while 
the 
spectators passed by. This leads to the excess of the average multiplicity 
in pp/\app\ collisions compared to that in \ep\ interactions at 
$\Ecme=\Ecmp/3$ as it is  observed from  Fig.~\ref{fig:multshe}.

 Comparing further 
the average 
 multiplicity values from pp/\app\ collisions to
those from heavy-ion reactions,  one finds that the 
data points  are amazingly close 
 to each other when the heavy-ion data is confronted with the hadronic 
data 
at the energy 
$\Ecmp  =3\, \Ecmn$ and, according to the just made observation,  
 to the \ep\ data  at the same energy as the nucleus-nucleus 
reaction data are taken.
 All these findings  agree well with our interpretation of the 
multihadron production process and the consequences the bulk variables to 
  be similar in the reactions considered when measured at $\Ecmn$ and 
$\Ecme$ being about a third of $ \Ecmp$. 
The inclusion of the tripling energy factor 
indeed allows {\it  energy-independent}
description of  
such a fundamental variable as the mean 
multiplicity  
in  
{\it simultaneously} 
\ep, pp/\app\ and central nucleus-nucleus collisions. 
This shows that the  particle production process in  headon 
nucleus-nucleus collisions 
 is 
derived by the energy deposited in the Lorentz-contracted volume by 
a single pair of effectively 
structureless nucleons similar to that 
 in \ep\ annihilation and of quark pair interaction in pp/\app\ 
collisions.   
 To note is that an examination of  Fig.~\ref{fig:multshe} reveals 
that there is no need to rescale  
the $\Ecme$ by a factor 
of 1/2 to match the \app\ data as earlier was assumed   
\ct{ph-sim,st02,busza}, 
while
  recognised \ct{ph-rev} to  unreasonably shift the \ep\ data\footnote{
Recall that the factor 1/3 but not 1/2 was already found earlier in 
\ct{ee3pp} to 
give an 
appropriate 
rescaling of the pp mean multiplicity data relative to those from \ep\ 
annihilation. 
}  
 on the 
pseudorapidity density at 
midrapidity when compared to the heavy-ion measurements. This 
discrepancy, as shown,  
 finds its explanation in our consideration, within which the 
data on $\it 
both$ the mean 
multiplicity and the midrapidity density ({\sl vide infra}) are 
self-consistently
 matched for different reactions.  

  Fig.~\ref{fig:multshe} shows  that 
the mean multiplicities measured in  different types of interactions  
are close to each other starting from the SPS $\Ecmn$ energies, and 
become particularly close  
at 
$\Ecmn \gtsim$~50~GeV.  At lower energies, however,  small
deviation in the measurements is visible. The nucleus-nucleus  data points
are slightly 
below  the measurements from \ep\ and hadronic experiments. At $\Ecmn 
\lssim$~10~GeV, the nuclear data  departure 
increases further, and the data 
start to decrease faster with the 
energy decreasing than the measurements from pp and \ep\ 
experiments do.   On the 
other hand, as c.m.
energy increases above a few tens GeV, the heavy-ion data start to exceed 
the \ep\ data and 
reaches the mean multiplicity values from \app\ interactions. 
 From this, one concludes on the two different regions of the rise of the 
mean 
multiplicity nuclear data with the energy increase: one being steeper at 
$\Ecmn$ of a few tens GeV, and another one  slower at lower energies down 
to a few GeV. One also can see that 
the multiplicities in nuclear 
reactions increase faster with the c.m. energy than this can be  
found in other interactions, while the effect is relatively small 
keeping the data points close to each other at fixed c.m. energy 
$\Ecmn$ and $\Ecme$ of  the value of $\Ecmp/3$.

The observations made can be understood in terms of the overlap zone. 
 At relatively low energies, 
the initial thermalized state, formed in headon
nuclear collisions, 
are most likely to occur in the overlap zone
of colliding nuclei along the incident direction, so the  nucleons at the 
periphery are  not contributing.
 Since the energy of the collision is 
small and, therefore,  is almost entirely and fast converted into 
particles produced, 
 the 
rest parts of the nuclei start to  fragment into pieces by expanding 
overlap zone. The 
smaller the energy is, the less particles will be produced in a collision. 
 At low energies, due to our approach, the mean multiplicity in 
pp collisions at $\Ecmp$ 
is expected  
to be 
larger than in nucleus-nucleus collision at $\Ecmn=\Ecmp/3$. As 
the $\Ecmn$  
increases,  the mean multiplicity in nucleus-nucleus collisions is 
supposed 
to increase 
faster than that in pp/\app\ interactions at $\Ecmp=3\Ecmn$.
Indeed, in pp/\app\  collision we consider a single pair of constituent 
quarks to 
form participants, while in nuclear collisions, the 
effectively structureless protons entirely participate. So, at the same 
energy per participant,  
at low energies, in (anti)proton collisions, all 
the energy of a pair of interacting constituent quarks  
is converted into 
particles produced, while in nuclear collision some fraction of energy 
is used  to  break-up  nuclei by expansion of the small overlap zone. 
 As the energy increases, 
the  nature of the rise in (anti)proton collisions should not  change 
significantly  since (anti)protons are considered interacting the same 
way as at lower energies, \ie\ \va\ a pair of 
constituent quarks depositing their energy into secondaries. In contrary,
 in (central) nuclear collisions, the Lorentz contraction starts to play 
a  significant role  involving more nucleons into overlap zone. 
 Due to this,  more energy becomes available for particle 
production which makes 
the mean 
multiplicity  rise faster than 
at lower energies.   
 To add is that this effect attracts currently special consideration 
in different 
approaches \ct{low-energy}.

Note that the total multiplicity is not very sensitive to 
the above, and this is seen in Fig.~\ref{fig:multshe}, where, as we 
already 
mentioned, the difference between hadronic and nuclear data 
is small, once tripling energy factor is taken into account. 
 On the other hand, due to fact that differences are related mostly to 
the 
interaction zone and consequently, to the central rapidity region, one 
would expect 
the differences to be more pronounced in the midrapidity comparison, as 
we indeed find below.
 \medskip

\begin{figure}[t]
 \vspace*{-.9cm}
\epsfysize=13cm
  \begin{center} 
\epsffile{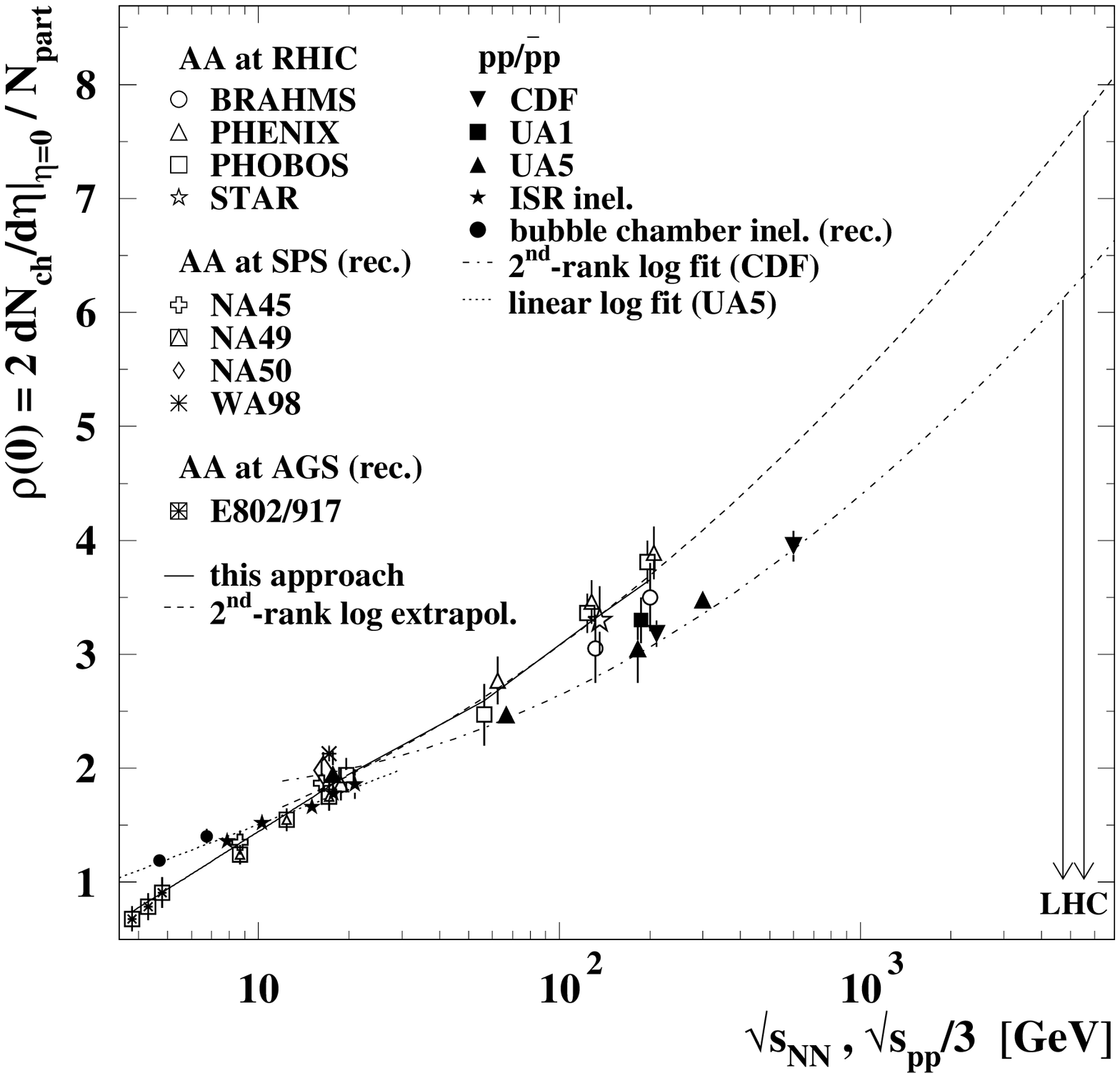}
  \end{center} 
 \vs{-.7cm}
\caption{   
Pseudorapidity density of charged particles per participant pair at 
midrapidity as a function 
of the c.m. energy of 
collision. The open and combined symbols show the pseudorapidity 
density values per 
participant pair  
\vrs\ c.m. energy per nucleon, $\Ecmn$, measured in the most central 
heavy-ion collisions 
at RHIC by BRAHMS \ct{br-200} 
 ($\scriptscriptstyle \bigcirc$), 
PHENIX \ct{phx-syststuds}
($\scriptstyle \triangle$), 
PHOBOS 
\ct{ph-56130,ph-colgeom} 
 ($\scriptstyle \square$), 
and STAR 
\ct{star-200} 
 ($\star$) 
  Collaborations,  
and the density values recalculated in \ct{phx-syststuds} from the 
measurements taken at CERN 
SPS by 
CERES/NA45 \ct{na45} (+), NA49 \ct{na49} ($\triangleinsquare$),  
NA50  \ct{na50}  
 ($\scriptstyle \diamondsuit$) 
and WA98 \ct{wa98}
 ($\timesplus$) 
Collaborations and at Fermilab AGS by 
E802 and E917 Collaborations \ct{ags} ($\timesplussquared$).
 The nuclear data at $\Ecmn$ around 20~GeV and the RHIC data at 
$\Ecmn=130$~GeV and 200~GeV are given spread horizontally 
for clarity. 
The PHENIX data at $\Ecmn=62.4$~GeV is taken from \ct{phx-milov}.
 The solid symbols show the pseudorapidity density values \vrs\ c.m. 
energy 
$\Ecmp/3$ as 
measured 
in non-single diffractive ${\bar {\rm p}}{\rm p}$ collisions 
 by UA1 \ct{ua1} 
 ($\scriptstyle \blacksquare$) 
and UA5  \ct{ua5-53900,ua5-546} 
 ($\blacktriangle$) 
 Collaborations at CERN SPS,  
by UA5 at CERN ISR 
($\Ecmp=$~53~GeV), by CDF \col\ at Fermilab
\ct{cdf}  
 ($\blacktriangledown$), 
and in inelastic pp collisions from the ISR \ct{isr-thome} ($\scriptstyle 
\bigstar$) and bubble chamber \ct{fnalmult,fnal-rap205} 
($\bullet$) experiments.
The data from the bubble  chamber experiments \ct{fnalmult,fnal-rap205} 
are given 
as recalculated in \ct{ua5-53900}.  
 The solid  line connects the predictions
 from Eq.~(\ref{prap0}). The dashed line 
gives the fit to the calculations using the 2nd order 
log-polynomial fit 
function analogous to that
 used \ct{cdf}
 in
 \app\ data. 
 The
 fit function from \ct{cdf} is shown by the dashed-dotted line. The dotted 
line shows the linear 
log 
 approximation
of UA5 to inelastic events \ct{ua5-53900}.   
 The arrows show the expectations for the LHC.
Note that \ep\ data  at $\Ecme=$~14~GeV to 200~GeV (not shown) 
follows 
 the heavy-ion data  \ct{ph-rev}.
 }
\label{fig:rap0}
\end{figure}

{\bf 4.} 
In Fig.~\ref{fig:rap0}, 
we compare 
the values of the
pseudorapidity densities per participant pair at midrapidity as  
 a function of c.m. 
energy, 
 from
most central nucleus-nucleus collisions at RHIC 
\ct{ph-56130,phx-syststuds,ph-colgeom,br-200,star-200}, CERN SPS 
\ct{na45,na49,na50,wa98}, and AGS \ct{ags}  
 \vrs\ those measured in 
pp/\app\ 
interactions at CERN \ct{ua5-53900,ua5-546,ua1,isr-thome} and Fermilab 
\ct{cdf,fnalmult,fnal-rap205}. The comparison is 
 given the same way as the mean  multiplicity is shown in 
Fig.~\ref{fig:multshe}, 
\ie\ the data from nucleus-nucleus collisions are plotted at the energy 
$\Ecmn=\Ecmp/3$. One can again see that up to the existing $\Ecmn$, the 
 data  from hadronic and 
nuclear experiments are close to each other being consistent with our
 interpretation. The measurements from the two types of collisions 
coincide at  8~GeV~$<\Ecmn<20$~GeV and 
 are of the magnitude of the spread of heavy-ion data 
points
at the highest  energy of 200~GeV. For a few GeV energy shown, one  can 
also 
see the visible difference.  
 The data shown in Fig.~\ref{fig:rap0} indicates that 
 the deviation between 
 the two types of collisions
increases with the c.m. energy due to  faster increase of the 
midrapidity density values obtained in heavy-ion collisions in comparison 
with  
those measured in \app\ interactions. At lower energies too, the nuclear 
data, being lower than the pp data, increases faster. 
   The latter means that,  as we discussed above,
in contrast to the
 mean multiplicity,
which is in general 
defined by the total yield of the reaction,  so being less sensitive to 
reaction details, 
the
midrapidity density depends on some additional factor. 
As the midrapidity density 
is measured in the very central region, where the
participants longitudinal velocities are zeroed, 
it is natural 
to assume 
that this factor is related 
to
the size  of the Lorentz-contracted volume of the initial thermalized
system determined by  participating patterns.

To take into account  the corresponding correction, let us 
consider our 
picture in the 
framework of the   
Landau model which reasonably well describes the bulk variables measured 
and is, by its nature, near to our interpretation as discussed above.
 Using this model, one finds  for the ratio of the charged particle
rapidity density $\rho(y)=(2/N_{\rm part})dN_{\rm ch}/dy$ per participant 
pair 
at the 
midrapidity value $y=0$ 
in 
heavy-ion reaction, $\rho_{\rm NN}$, to the density $\rho_{\rm pp}$ 
in 
pp/\app\ interaction,
 \begin{equation}
\frac{\rho_{\rm NN}(0)}{\rho_{\rm pp}(0)}=
  \frac{2\,N_{\rm ch}}{N_{\rm part}\, N^{\rm pp}_{\rm ch}}
 \, 
\sqrt{\frac{L_{\rm pp}}{L_{\rm NN}}}\,. 
\label{rap0}
\end{equation}

\nopar
Here,  $N_{\rm part}$ is the number of participants in 
heavy-ion collision, 
$N_{\rm ch}$ ($N_{\rm ch}^{\rm pp}$) 
 is the 
multiplicity in nucleus-nucleus (pp/\app) collision 
and $L= \ln \frac {\sqrt {s}}{2m}$, where $m$ is the mass of a
participating 
pattern, \eg\ of a proton, $m_{\rm p}$, in central heavy-ion collisions.  
 According to our interpretation, we compare in the ratio (\ref{rap0}) the 
rapidity 
density  $\rho_{\rm NN}(0)$
at $\Ecmn$ to the rapidity density 
  $\rho_{\rm pp}(0)$
at $\Ecmp/3$. Due to the above, we consider a 
constituent quark of mass $\frac{1}{3}m_{\rm p}$ as a participating 
pattern in 
pp/\app\ 
collisions, and a proton as an effectively structureless participant in  
most central nucleus-nucleus 
collisions.
 Then, from Eq.~(\ref{rap0}) one  obtains:
 \begin{equation}
  \rho_{\rm NN}(0)= \rho_{\rm pp}(0) \,   
  \frac{2\,N_{\rm ch}}{N_{\rm part}\, N^{\rm pp}_{\rm ch}}
 \,
\sqrt{1-\frac{4 \ln 3}{\ln\, (4 m_{\rm p}^2/s_{\rm NN})} }\,. 
\label{prap0}
\end{equation}

Using the fact that the transformation factor from rapidity 
to pseudorapidity does not 
influence the above ratio and substituting the data values of $N_{\rm 
ch}$ and $N^{\rm pp}_{\rm ch}$ shown in  
Fig.~\ref{fig:multshe} and of $\rho_{\rm pp}(0)$
shown in Fig.~\ref{fig:rap0}, one obtains 
from Eq.~(\ref{prap0})
the 
values of pseudorapidity density 
in nucleus-nucleus 
collisions. These values are    
 displayed in 
Fig.~\ref{fig:rap0} 
by solid line. One can see that the 
correction made provides  a good agreement between the  
calculated 
$\rho_{\rm NN}(0)$ 
values  and the
data. 
This
justifies the above argued midrapidity multiplicity 
dependence on 
 participant type determining the volume of the initially thermalized 
system.  Eq.~(\ref{prap0}) shows also the importance of the correction, 
 in particular in case 
of the
Landau model,  for the 
participant 
 type
 to be introduced to properly estimate  the midrapidity 
density.
 This correction is in agreement with our description where the proper 
participants have to be considered for the corresponding reaction. 
 To note is that the pseudorapidity density at midrapidity in \ep\ 
annihilation is shown 
\ct{ph-rev} to 
 coincide with that from nucleus-nucleus data in the wide energy range and 
is in agreement with our 
assumptions and the above expectations. 

As we discuss above for the mean multiplicity, one can see that 
the pseudorapidity density at midrapidity is indeed  more sensitive to the 
differences in the multiparticle production process at lower and 
higher energies. The increase of the density measured in nuclear 
collisions 
at $\Ecmn$ less than the highest SPS energy  is steeper than it is above 
it. 
This is also true for pp/\app\ data which is fitted with different log 
functions: with the single log fit at lower $\Ecmp$ \ct{ua5-53900} and 
with 
2nd order log polynomial
 at higher energies \ct{cdf}. It is worth noting how well our 
consideration of tripling energy and properly treating the type of 
participants -- quarks in pp/\app\ and protons in nuclear collisions 
-- 
takes 
into account the different energy behaviour in pp/\app\ collisions from 
which  
the nuclear data can be deduced.     
 Note  that the same two regions we find here studying the 
pseudorapidity density at midrapidity $\rho_{\rm NN}(0)$, recently has 
been 
reported in \ct{phx-syststuds,phx-milov} by PHENIX, when analysing
the ratio of the transverse energy density to the pseudorapidity density 
at midrapidity as function of energy.
From this finding,  one can expect to obtain 
similarities in the transverse energy density in pp/\app\ \vrs\ that in 
headon heavy-ion collisions in the frame of our description.  
 The transition region at the SPS energies has been found also by  NA49 
\ct{na49-mult}, which can be treated within our approach too, without any 
additional assumptions.

From the above, we conclude that the description proposed here allows
 explaining the similarity obtained between the bulk variables measured in 
central
heavy ion collisions compared to other types of reactions. Namely, we 
consider the
constituent quark structure as a frame governing the formation of the
initial thermalized state in the collision zone. Then a pair of
structureless participating patterns being stopped in a Lorentz-contracted
volume is the main source of energy dissipating into particles produced.
 This makes only a fraction of the total energy of colliding
composite object like (anti)proton available for formation of
secondary particles. In contrary, in headon collision of nuclei, nucleons
are considered as being structureless, similar to \ep\
annihilation, owing to almost coherent deposition of energy by all
participating constituent quarks. This provides the total energy per 
nucleon being available for the formation of the thermalized
Lorentz-contracted volume and consequently, for the secondary particle
production. Therefore, in headon nuclear  collisions, and in \ep\ 
annihilation as 
well, a tripling 
factor is
needed to be taken into account for energy and mass of interacting
patterns to adjust the global variables 
 to those in pp/\app\ collisions. 
It is worth noting that recently, the constituent quark picture has been 
exploited to
reasonably
model the heavy-ion  pseudorapidity and transverse 
energy data \ct{ind}.
 \medskip

{\bf 5.}  An intriguing issue, which can be considered from the 
interpretation 
 given here in tracing similarities of  the 
bulk observables in nucleus-nucleus collisions and in ``simpler'' 
reactions,  particularly  in such like
nucleon-nucleon 
ones, is to make some predictions for the bulk variables in 
asymmetric  nucleon-nucleus collisions.
 In the frame of our consideration, in these interactions, the mean 
multiplicity value per pair of participant measured at some $\Ecmn$ is 
expected to have the same 
value as that is in pp/\app\ collisions at the same $\Ecmp$  as $\Ecmn$.
Indeed, 
assuming an   incident 
proton in p-nucleus collisions interacts in a way it would interact in 
pp collision,  the
secondary particles in the reaction are considered to be created 
 out  of energy deposited by interaction of a single pair of constituent 
quarks, one of which is from the proton and another one 
from a nucleon of the interacting nucleus. So, in fact, only this pair of 
constituent quarks converts their energy into secondary particles similar 
to that  in pp/\app\ interactions. 
 This  also implies  that the mean multiplicity being defined mostly 
by the 
energy 
deposited by participants -- by the pair of constituent quarks -- 
 is not expected to depend on 
 the centrality of nucleus-induced  
collisions, \ie\ on the 
 number $N_{\rm part}$ of participants (within uncertainties 
due to intranuclear effects, \eg\ Fermi motion). 
 These expectations are recently shown 
to be well confirmed in 
the RHIC data on deuteron-gold interactions 
at $\Ecmn=$~200~GeV \ct{ph-rev,ph-dAu-talk,ph-dAu}.  
 Moreover, the effect of the similarity obtained at RHIC is shown 
\ct{ph-rev,ph-dAu-talk,ph-dAu} to be 
true for hadron-nucleus collisions at $\Ecmn\approx $~10--20~GeV too. 
 The same seems to be correct also for the pseudorapidity density at 
midrapidity, which is already supported to be a trend \ct{ph-rev,ph-dAu}. 
This can also explain the necessity of the multiplication \ct{levin-errat} 
of the pp/\app\ 
pseudorapidity 
distribution 
by a number of the Au 
participants to correct the QCD saturation model predictions \ct{levin-dA}
to fit 
the data gold nucleus region of the 
pseudorapidity 
distribution
 from 
dAu interactions.
 \medskip 

{\bf 6.} 
  As it is found, see Fig.~\ref{fig:rap0}, the 
pseudorapidity 
density at midrapidity shows increase in  the  difference between the 
values obtained 
from  central nucleus-nucleus 
collisions and those from pp/\app\ interactions as the 
c.m. energy of reactions  
increases above the highest SPS energy. As shown above, this difference 
is 
connected with a size of the 
Lorentz-contracted volume.  
 At the c.m. energies available today in nucleus-nucleus experiments, this 
difference is  comparable with the  
difference in the measurements. It is of interest 
 to know the density values  for higher than the experimentally 
reached $\Ecmn$ in order to check 
 whether 
the 
deviation between the pp/\app\ and 
nucleus-nucleus midrapidity densities becomes more 
pronounced as it is expected.  Unfortunately, no measurements, 
except the midrapidity 
density for 
pp/\app\ data, are available for the
variables in Eq.~(\ref{prap0}) beyond $\Ecmn=$~200~GeV. 
The densities  for \app\ interactions are measured up to 
$\Ecmp=$~1.8~TeV, 
and these 
high-energy \app\ data
can be compared to at least  some estimates for the 
midrapidity density in nucleus-nucleus collisions at $\Ecmn=\Ecmp/3$.  
 To this end, we 
extrapolated the values calculated from 
Eq.~(\ref{prap0}) 
 utilizing 
the function analogous to  that found \ct{cdf} to fit well the 
\app\ 
data.  Both the prediction for nucleus-nucleus collisions and the fit for 
\app\ 
data are shown in Fig.~\ref{fig:rap0}.

The dependence of the
midrapidity density
on the
c.m. energy in nucleus-nucleus interactions, which we 
 show by the dashed line in Fig.~\ref{fig:rap0},  agrees well  
with the  calculations we made above based on Eq.~(\ref{prap0}) shown 
by 
solid 
line.  
 This dependence
 indicates that the 
 midrapidity density in heavy ions increases faster with the c.m. 
energy
than that in 
pp/\app\ collisions considered at the three times larger c.m. energy 
$\Ecmp$.  As we predict within our consideration and  
show in Fig.~\ref{fig:rap0} in an 
 assumption that the same behaviour 
as at the SPS--RHIC energies extends up to the LHC energies, 
the density $\rho_{\rm NN}(0)$ is found to be $\sim 7.7$ at  
the
c.m. energy $\Ecmn=$~5.5~TeV of heavy-ion collisions at LHC. From the CDF 
fit \ct{cdf} and assuming it covers LHC  energies, one obtains 
the  value of $\sim 6.1$ for the pseudorapidity density at midrapidity 
expected for  
 pp interactions at LHC at $\Ecmp=$~14~TeV.  
 The value we obtain for $\rho_{\rm NN}(0)$ at LHC energy is consistent
with
that of $\sim 6.1$ given in the PHENIX extrapolation 
\ct{phx-syststuds,phx-milov} within 
1-2 particle 
error acceptable in the calculations we made. Our result is in a good 
agreement with the recent  ATLAS Monte Carlo tuned values \ct{atlas}.
 Taking into account that the LHC $\Ecmn$ is close enough to $\Ecmp/3$,  
the nearness of the values, predicted for the LHC  by us and estimated 
independently and based on the experimental fit in the wide range of 
energies by 
PHENIX \ct{phx-syststuds,phx-milov}, demonstrates that our interpretation 
provides an 
experimentally 
 grounded description and predictive ability.
It is interesting to note that as the energy changes significantly the 
particle production process seems  not to change from that above highest 
SPS nuclear collision 
energies, in contrast to what happens when the $\Ecmn$ departs to the 
SPS energies from lower $\Ecmn$ as discussed.  

 As soon as we describe nucleus-nucleus midrapidity density using pp/\app\  
data in wide energy range spanning form a few GeV to hundreds GeV, now we 
can solve Eq.~(\ref{prap0}) for the mean multiplicity $N_{\rm 
ch}/(0.5 N_{\rm part})$ to 
describe the nuclear data where  no measurements exist and to predict 
the mean multiplicity energy 
dependence at higher energies.  In this calculations, we use the fits of 
$\rho_{\rm pp}(0)$ \ct{cdf} and $N^{\rm pp}_{\rm ch}$ \ct{ua5-53900} and 
approximation for
$\rho_{\rm NN}(0)$ 
as functions of the c.m. energy  shown in 
Figs.~\ref{fig:multshe} and
\ref{fig:rap0}. From the resulted curve for 
$N_{\rm ch}/(0.5 N_{\rm part})$ as a function of $\Ecmn$ plotted in 
Fig.~\ref{fig:multshe}, 
one finds that the heavy-ion 
mean multiplicity per participant pair at $\Ecmn=$~5.5~TeV is just about 
$10\%$ above the $N^{\rm pp}_{\rm ch}(\Ecmp)$ approximation at 
$\Ecmp=$~14 TeV and about 3.3 
times 
larger the data mean multiplicity from heavy-ion collisions at 
$\Ecmn=$~200~GeV. 
Again, this number is 
comparable with 
the estimate made from the pseudorapidity density spectra by PHENIX 
\ct{phx-syststuds,phx-milov} and points out to no evidence for change of 
behaviour  in the energy dependence 
  as the $\Ecmn$  increases by about two 
magnitudes from  
 the top SPS energy.      
 Nevertheless, one can see that the data obtained at the highest  RHIC 
energy  give  a hint to 
some 
border-like behaviour of the mean multiplicity where the 
pp/\app\ 
 data saturate the nuclear data, and another region of the rise 
is possible to be 
found (as at 
low energies). This  makes  
heavy-ion 
experiments 
at $\Ecmn>200$~GeV of particular interest.

Given the importance of the multiplicity and produced particle spectra as
 control observables of the particle production process, 
various theoretical descriptions have been proposed 
and 
confronted  to the nucleus-nucleus collision data,  in 
  particular  to the mean multiplicity and
the midrapidity density measurements \ct{ph-rev,phx-rev,br-rev,theory},  
the bulk variables considered 
here.
  Most of the models are shown to reproduce reasonably at least some part 
of the
measurements or the trend obtained in the data, while they are found to be 
less 
successful 
in describing all the experimental findings. Despite many questions are 
still open and there are significant challenges for theory, the 
message of 
 inability to reproduce the measurements taken at highest RHIC 
energies by scenarios solely based on interactions between hadronic 
objects is 
clearly given. The system produced at highest density and  temperature 
ever reached experimentally requires a partonic approach to be explored 
for its description. 
In this sense, our interpretation invokes partonic picture  
being the necessary ingredient as  the description utilizes the 
constituent quark framework.
 \medskip 

{\bf 7.} 
  In summary, we 
analyse  
the similarities of the bulk observables obtained in heavy-ion 
reactions,  (anti)proton-proton and \ep\ interactions
in the large c.m. energy interval from a few GeV to hundreds GeV. 
 The similarities of such bulk 
observables like the mean multiplicity and pseudorapidity density at 
midrapidity are compared to our expectations under 
assumption of  
the 
universality of a mechanism of the multiparticle production  
 in different 
types of  
interactions.
 Within the description proposed, secondary particles are 
produced  according to the amount of energy deposited by participants 
and 
depending on their type.  The bulk observables in headon 
nuclear collisions are treated on the base of  interactions of 
nucleons, 
considered 
as effectively structureless patterns wholly depositing their energy, 
similar to that in \ep\ annihilation. 
 It is shown that the observables obtained from  
nucleon-nucleon 
collisions are similar to those from nuclear or \ep\ interactions when one 
considers the single constituent quark pair interaction picture for 
nucleon-nucleon 
collisions at the 
c.m. energy three times 
larger  the energy of nuclear or \ep\ collisions. The correction to 
take into account the tripling factor in the c.m. energy and for 
proper treatment of participating patterns is found to explain the actual 
measurements.  Two separated regions in the measurements   
with a 
boundary about highest SPS nuclear experiment energy are reproduced.
The approach is extended to and the consequences are made 
for particle-nucleus collisions showing  agreement with recent RHIC data. 
The predictions 
for higher LHC energies within the proposed interpretation are 
made, found to follow the experimentally obtained extrapolations. 
 As shown, experiments at RHIC, 
given a remarkable opportunity to expand investigations 
to 
the 
highest so far achieved density of matter,  
 allow
to reveal
signatures of new phenomena and provide 
indispensable knowledge on the dynamics of
strong 
interactions. This makes of great interest further comprehensive 
analyses of the data 
obtained up to RHIC energies and to be obtained at the   
higher energies challenging to the better
understanding of the discoveries made and 
 testing the predictions.  
\bigskip
\medskip

\nopar
{\bf \Large Acknowledgements}\\

\nopar
We are indebted to I.M. Dremin, J. Ellis, W. Kittel, A. Milov, W. Ochs, D. 
Plane,
V.N. Roinishvili, and P. Steinberg 
for 
fruitful 
discussion 
and 
helpful 
comments.


\end{document}